\documentclass[aps,prb,preprint,superscriptaddress]{revtex4-1}

\usepackage{graphicx} 
\usepackage{graphics} 
\usepackage{amsfonts} 
\usepackage{amssymb} %
\usepackage{color} 
\usepackage{epstopdf}
\usepackage{hyperref}
\hypersetup{colorlinks = true, allcolors = {blue}, allbordercolors = {1 1 1}}

\begin{document}

\title{Observation of the Nonlinear Phase Shift Due to Single Post-Selected Photons}

\author{Amir \surname{Feizpour}}
\author{Matin \surname{Hallaji}}
\author{Greg \surname{Dmochowski}}
 \affiliation{Centre for Quantum Information and Quantum Control and Institute for Optical Sciences, Department
of Physics, University of Toronto, 60 St. George Street, Toronto, Ontario,
Canada M5S1A7}
\author{Aephraim M. \surname{Steinberg}}
 \affiliation{Centre for Quantum Information and Quantum Control and Institute for Optical Sciences, Department
of Physics, University of Toronto, 60 St. George Street, Toronto, Ontario,
Canada M5S1A7}
\affiliation{Canadian Institute For Advanced Research, 180 Dundas St. W., Toronto Ontario, Canada M5G1Z8}

\maketitle

\textbf{
Over the past years, there have been many efforts towards generating interactions between two optical beams so strong that they could be observed at the level of individual photons. 
Such strong interactions, beyond opening up a new regime in optics, could lead to technologies such as all-optical quantum information processing. 
However, the extreme weakness of photon-photon scattering has hindered any attempt to observe such interactions at the level of single particles. 
Here we implement a strong optical nonlinearity using electromagnetically-induced transparency and slow light, and directly measure the resulting nonlinear phase shift for individual photons. 
This is done by illuminating the sample with a weak classical pulse with as few as 0.5 photons per pulse on average, and using post-selection to determine whether a given pulse contained (approximately) 0 or 1 photons. 
We present clear data showing the quantized dependence of a probe beam's measured phase shift on the post-selection result, for a range of input pulse intensities. 
We believe that this represents the first direct measurement of the cross-phase shift due to single photons.}

Modern optical physics has revolved principally about two poles: nonlinear optics, where rich effects are generated through the interactions of photons with one another, but which because of the weakness of those interactions typically manifests itself only for pulses containing billions of photons; 
and quantum optics, where phenomena such as entanglement have been widely studied, but where photon-photon interactions are negligible.  
In fact, it has been well known in quantum optics since the Nobel Prize-winning work of Roy Glauber that in the linear regime, the classical and quantum theories of electromagnetism make identical predictions (except in the presence of nonclassical sources of light - which themselves rely on nonlinear effects) \cite{glauber1963coherent,glauber1963quantum}. 
It has therefore long been a dream to move into the realm of ``quantum nonlinear optics," where sufficiently strong interactions could create a complex many-body interacting quantum system in the context of optics. 
Such interactions could enable the production and detection of novel entangled states including few-photon bound states \cite{deutsch1992diphotons,firstenberg2013attractive}, as well as new architectures for non-demolition measurement of photon number \cite{ImotoQND,braginsky_quantum_1980,grangier_quantum_1998}, quantum teleportation \cite{vitali2000complete}, low light level switching \cite{HarrisSwitching1998}, and quantum logic gates \cite{milburn1989fredkin}.  
Decades ago, important steps in this direction were taken in the realm of cavity QED \cite{kimble1995cPhase}, leading for instance to major advances in the study of entanglement \cite{raimond_manipulating_2001,guerlin_progressive_2007,gleyzes_quantum_2007,rauschenbeutel_step-by-step_2000}, and more recent dramatic developments in the context of superconducting qubits \cite{wallraff2004strong,devoret2013superconducting}. 
Greatly enhanced interactions for photons have recently been observed, using electromagnetically induced transparency and slow light \cite{schmidt1996giant}, microstructured fibres \cite{matsuda2009observation}, atoms in hollow fibres \cite{venkataraman2013phase}, a single atom strongly coupled to a micro-resonator cavity \cite{PhysRevLett.111.193601}, atoms or other impurities near tapered fibres \cite{hendrickson2009nonlinear,spillane2008observation} or bottle resonators \cite{volz_nonlinear_2014}.
Most recently, the application of ``Rydberg blockades'' \cite{urban2009observation,saffman_quantum_2010} has enabled huge interaction strengths, already leading to the observation of strongly modified quantum statistics \cite{parigi_observation_2012,peyronel2012quantum,firstenberg2013attractive,Chen16082013,PhysRevLett.112.073901}.  
Sum-frequency generation of two heralded single photons has also been demonstrated \cite{guerreiro_nonlinear_2014}.
In parallel, there has been some controversy about the applicability of such strong interactions to quantum logic due to fundamental noise limits \cite{shapiroPRA06single,shapiroNJP07weak,GeaBanaclochePRA10impossiblity} and suggestions that intermediate-strength nonlinearities might be a powerful approach \cite{nemotoPRL04nearly,munroNJP05weak}.  
At the present time, the nonlinear phase shift written by a single photon on a probe
beam has to our knowledge never been reported, and prior experiments have been performed with classical pulses with average photon numbers on the order of a few hundred \cite{loPRAattojoulePhysRevA.83.041804} in free space, or as low as 16 or 0.1 in a hollow-core fibre filled with atomic vapour \cite{venkataraman2013phase} or a nonlinear photonic-crystal fibre \cite{matsuda2009observation}, respectively.  

Here we demonstrate that by illuminating a sample of atoms with a weak coherent state, but post-selecting on subsequent detection of a photon at the far side of the sample, we can observe the nonlinear effect of that one additional photon on a probe beam. 
Similar ``intensity-field correlations" were previously used to enhance quantum effects in a cavity-QED system \cite{smith_capture_2002,foster_third-order_2002,foster_time-dependent_2002}.
Using electromagnetically induced transparency in a gas of laser-cooled $^{85}$Rb atoms, we report the observation of a nonlinear phase shift of 18 microradians per photon, and unequivocably demonstrate the contribution due to individual post-selected photons.  
The sample was illuminated with weak classical pulses with as few as 0.5 photons on average, and post-selection allowed us to subsequently determine whether a given pulse contained (approximately) 0 or 1 photons. 
We present clear data showing the quantized dependence of the measured phase shift on the post-selection result, for a range of input pulse intensities. 
This represents the first direct measurement of the cross-phase shift due to a single photon.

We use an atomic level scheme\cite{schmidt1996giant} based on electromagnetically-induced transparency\cite{fleischhauer2005electromagnetically} (EIT), which allows for very strong near-resonant interactions.
EIT is a coherent atomic effect in which different excitation pathways interfere destructively, eliminating linear absorption and simultaneously producing sharp dispersive slopes; see the inset of figure \ref{fig_setup}.
In order to establish EIT, two phase-coherent laser fields (probe and coupling) form a $\Lambda$-system, addressing a common excited state.
When the two-photon resonance condition is satisfied, i.e. their respective detunings are equal, the medium becomes transparent to these fields and each experiences a modified refractive index profile.
The presence of an additional `signal' field inside the medium serves to ac-Stark shift one of the ground states of the $\Lambda$-system, pulling the probe and coupling fields out of two-photon resonance. 
As a result, the probe field experiences a change in refractive index, acquiring a phase shift that is proportional to the slope of this steep dispersive feature.
For low signal power, the induced Stark shift is linear in signal intensity; 
the resulting probe phase shift is, therefore, proportional to the number of photons in the signal pulse.
Measurement of this phase shift provides information about the number of signal photons present in the interaction region.

Figure \ref{fig_setup} shows a schematic of the experimental setup as well as the level scheme.
A cloud of laser-cooled $^{85}$Rb atoms serves as the nonlinear optical medium, with the probe and coupling beams passing through orthogonal to each other.
Short signal pulses are sent counter-propagating to the probe and both beams are collected by highly reflective beam-splitters after the interaction.
A single-photon counting module (SPCM) is used to detect the signal pulses while the probe beam is detected by a fast avalanche photodiode (APD), operating in the linear regime;
see the `methods' section for the details of the probe phase measurement.

\begin{figure}[h]
  \centering %
  \includegraphics[width = \columnwidth]{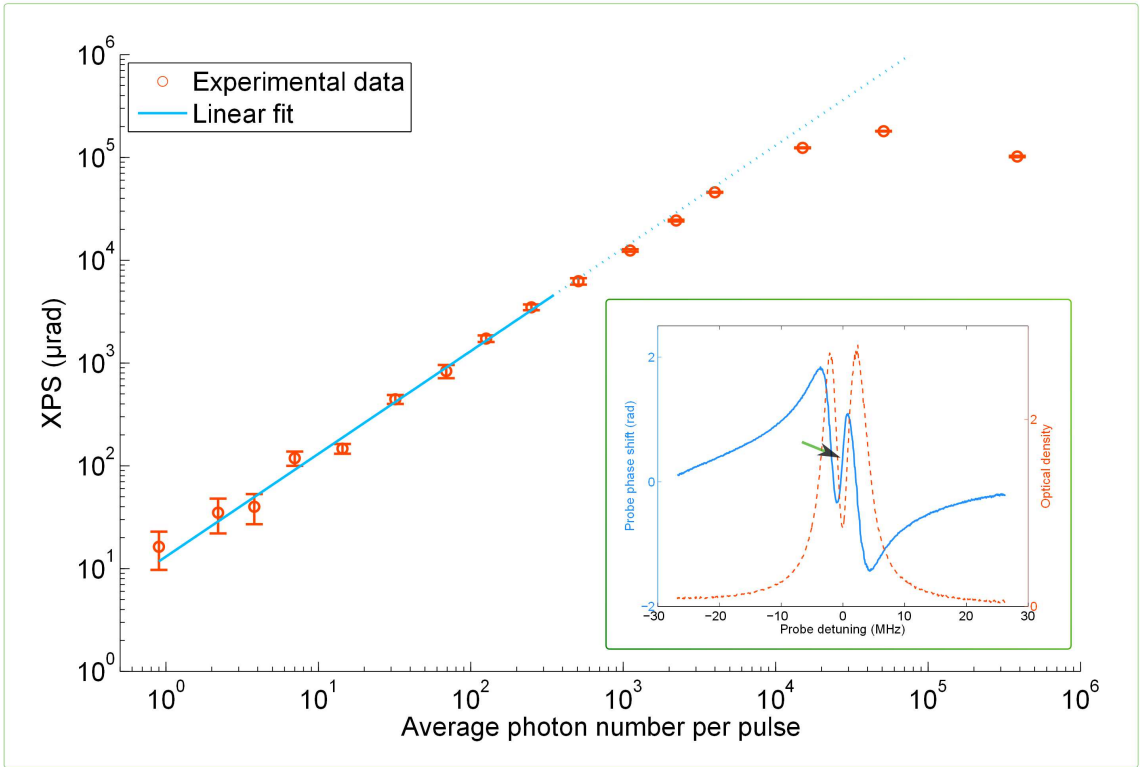}
  \caption{XPS versus average photon number per pulse.
  \label{fig_vsPhotonNumber}}
\end{figure}

We first measure the nonlinear cross-phase shift (XPS) for a range of signal pulse energies in order to determine the size of the per-photon effect.
Figure \ref{fig_vsPhotonNumber} plots the cross-phase shift on the probe field, versus average number of photons per signal pulse. 
The phase shift grows linearly for low photon numbers and a fit yields a slope of $13 \pm 1\ \mu$rad per photon. 
There is a saturation at high photon numbers that occurs when the induced ac-Stark shift becomes comparable to or larger than the half-width of the EIT window.
The lowest energy per pulse that we use here corresponds to an average of one photon per pulse, which is the lowest pulse energy ever used for cross-phase modulation in free space.




Although in the measurement described above we have observed the XPS due to signal pulses with an average photon number of 1, approximately $40\%$ of the pulses contain no photons at all; and about $25\%$ contain multiple photons. 
To observe the quantized effect of individual photons, we send in even weaker pulses - containing only one-half a photon on average - and trigger on subsequent detection of a single photon at an SPCM. 
These so-called `click' events occur only a small fraction of the time, due to the weak signal pulses as well as detector inefficiency.  
Because of this finite efficiency, the absence of a click does not preclude the possibility that there had, in fact, been a photon in the interaction region.  
Nevertheless - and neglecting for the moment background counts and multi-photon events - it can be shown that the best estimate possible of the photon number in the interaction region increases by exactly one when the detector fires, regardless of its efficiency.  
An intuitive picture for this involves recognizing that for an incident coherent state, all the ``undetected modes" are uncorrelated with the ``detected mode": gaining the information that there is 1 photon, rather than 0, in the detected mode thus has no effect on how many photons one should estimate are present in all other modes.  
One can therefore observe the effect of one additional photon on the probe beam by comparing cases when the SPCM fires with cases where it does not.

\begin{figure}[h]
  \centering %
  \includegraphics[width = \columnwidth]{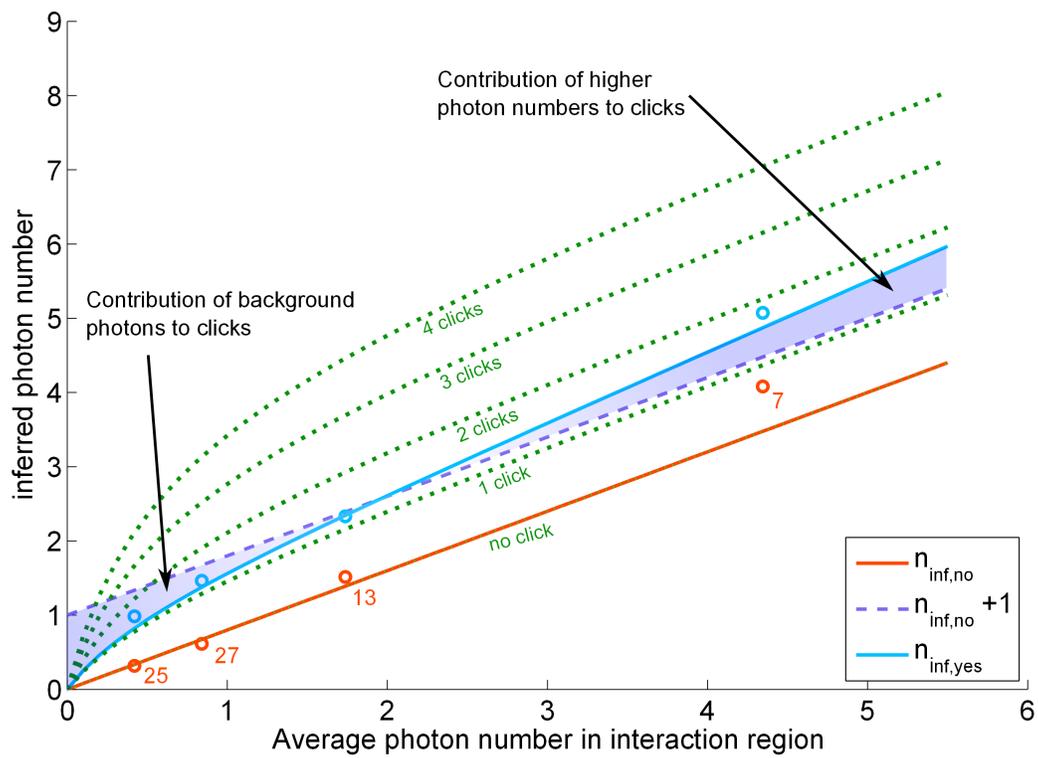}
  \caption{Inferred ($n_{inf}$) versus average photon number in the interaction region. 
  \label{fig_infPhNum}}
\end{figure}

When background and multi-photon events are included, one can still calculate the shift in 
inferred average photon number due to a `click' event, and it remains close to (but no longer exactly
equal to) 1; 
see the Supplementary Information for details.
This difference depends on the incident average photon number $|\alpha|^2$, the overall collection and detection efficiency $\eta$, and the background photon detection probability $P_b$.  
When the probability of clicks due to signal photons is much higher than that of ones due to background photons - but still much lower than 1 ($P_b \ll \eta|\alpha|^2 \ll 1$) - the difference is close to 1. 
However, at low incident photon numbers false clicks due to the background photons cause the inferred difference to be smaller than 1.
On the other hand, the difference can be larger than 1 when the contribution of higher photon numbers to the click events has to be included. 
Figure \ref{fig_infPhNum} plots the inferred average photon number in the interaction region, $n_{inf}$, versus the average incident photon number, $|\alpha|^2$, for click and no-click events. 
It also shows the photon number one would infer from a given number of counts at a photon-number resolving detector, for comparison.

\begin{figure}[h]
  \centering %
  \includegraphics[width = \columnwidth]{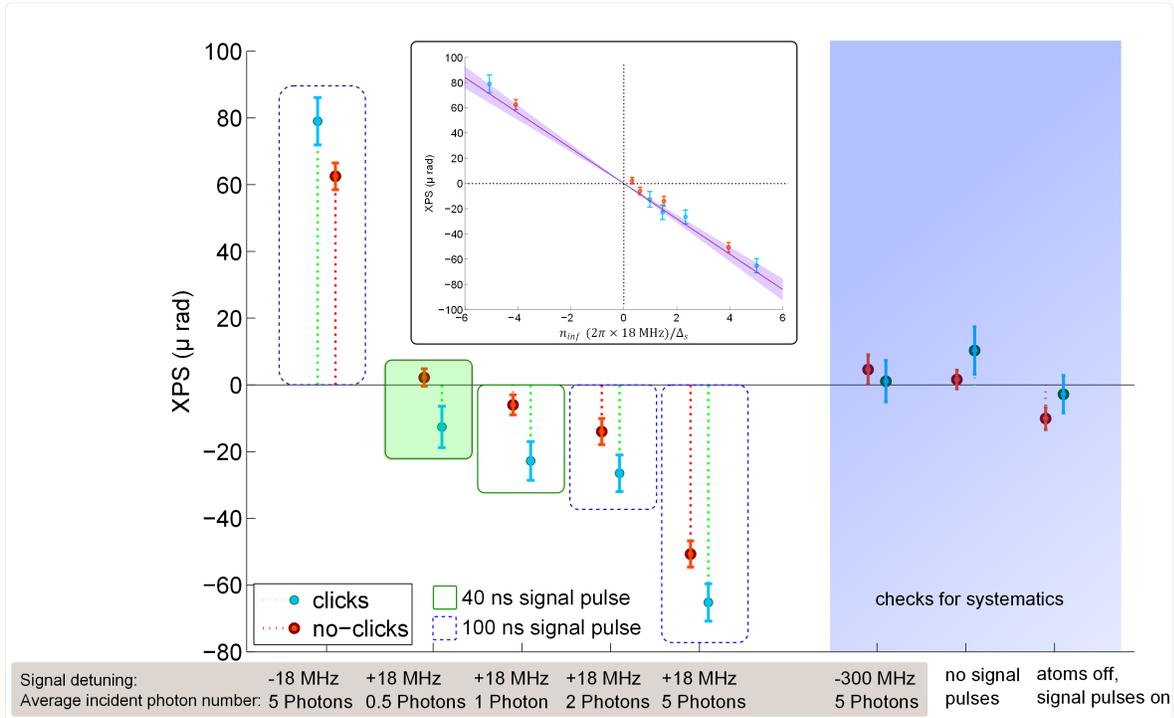}
  \caption{Post-selected XPS. 
  \label{fig_PSxps}}
\end{figure}

We can separate out the instances in which the SPCM detected a photon (the `click' cases) from those in which no click was registered by the detector.
This allows us to measure the nonlinear phase shift of the probe in the two cases separately, see figure \ref{fig_PSxps}.
For $0.5$ incident signal photons per pulse, the inferred average photon number in the interaction region for no-click events is roughly 0.3 whereas it is approximately 1 for click events; see the square shaded in green.
The phase shift measured for the no-click cases is statistically indistinguishable from zero ($2 \pm 3 \; \mu$rad), while the click events result in a non-zero phase shift of $-13\pm 6\;\mu$rad, consistent with the per-photon nonlinear phase shift of $-13\pm 1\;\mu$rad inferred from the fit to the data in figure \ref{fig_vsPhotonNumber}.

For the remaining data points, the average incident signal photon number and/or the center detuning of signal pulses was varied.
The most significant feature is that the nonlinear phase shifts for click events are always larger in absolute size than those for the no-click cases. 
The magnitude of the phase difference between click and no-click events averaged over all data points, i.e. the nonlinear phase shift due to a post-selected single photon, is $-15 \pm 3\ \mu$rad ($-18 \pm 4\ \mu$rad after correcting for finite background; see the Supplementary Information for details). 

In order to confirm that the observed effect is not due to systematics, we have also taken data in the absence of signal pulses or atoms, and with large signal detunings.
The data in the region shaded in blue displays these checks for systematics.
The most important feature to highlight is that the click- and no-click probe phase shifts for all systematic checks are equal to within the error-bars.

Finally, the inset of figure \ref{fig_PSxps} plots all the post-selected data versus inferred average photon number corrected for the variable signal detuning; see the Supplementary Information.
The solid line has a slope of $-14 \pm 1\ \mu$rad per photon which is inferred from the fit in figure \ref{fig_vsPhotonNumber}, the optical density, and the detuning dependence of the nonlinear effect. 
We see excellent agreement between the data presented here and the value of XPS per photon extracted from figure \ref{fig_vsPhotonNumber}.

The theoretical prediction for the magnitude of the cross-phase shift, given the parameters of our experiment, is 13 $\mu$rad based on the model presented in \cite{feizpour2014short};
see Supplementary Information for more details.
The values we measured in our experiment are in very good agreement to the theoretical value.

In summary, we have used EIT to make the first observation of cross-phase shift due to free-space signal pulses containing 1 or fewer photons on average.  
Moreover, by conditioning on detection of a photon in the signal pulse after it is transmitted, we have been able to detect the (quantized, in the ideal case) effect of a single additional photon on a probe beam, and measured this single-photon phase shift to be $-18 \pm 4\ \mu$rad, consistent with the results for classical signals.  
This is a step towards  the development of further "quantum nonlinear optics" techniques, including the potential for scalable quantum logic gates.

\newpage
\textbf{Figure captions}

\emph{Figure \ref{fig_setup}}.
Schematic of the experimental setup. 
Counter-propagating probe and signal beams are focused to a waist of $13\pm1\mu$m inside a cloud of laser cooled $^{85}$Rb atoms confined in a magneto-optical trap.
High reflectivity beam-splitters (90\% reflectivity) are used to collect the signal and probe beams after the interaction. 
A collimated coupling beam, propagating perpendicular to both probe and signal beams, creates a 2 MHz EIT window for the probe (see the inset).
The amplitude and phase of the probe are measured using frequency-domain interferometry.
The single-photon detection events on the SPCM are registered as tags in the amplitude of the probe. 
The inset shows the level scheme used. 

\emph{Figure \ref{fig_vsPhotonNumber}}.
XPS versus average photon number per pulse.
  The nonlinear phase shift depends linearly on the photon number at lower intensities. 
  A fit to the low-photon-number data yields a slope of $13 \pm 1\ \mu$rad per photon while the deviation at higher photon numbers arises due to higher-order nonlinearities.
  The inset shows a typical linear phase profile (green) and optical density (red) as seen by the probe with the arrow indicating where the on-resonance component of the probe laser is locked. 
  Other relevant parameters include signal center detuning = $-10$ MHz, OD = 2, EIT widow width = 2 MHz. 
  
\emph{Figure \ref{fig_infPhNum}}.
Inferred ($n_{inf}$) versus average photon number in the interaction region.
  The overall collection efficiency is assumed to be $20\%$ and the background click rate is taken to be $10\%$ for solid and dotted lines.
  The circles show the photon number values inferred for the data points in figure \ref{fig_PSxps} for no-click (red) and click (blue) events.  
  The overall efficiency percentage for each data point (numbers beside circles) is slightly different which accounts for the discrepancies between the data points and the solid curves.  
  The average photon number in the interaction region for the data points is lower than the incident photon number because of the finite signal absorption.
  The dotted green lines show the photon number which would be inferred were a number-resolving detector used.
  The solid blue line could also be obtained from a weighted average of the dotted lines with non-zero number of clicks.

\emph{Figure \ref{fig_PSxps}}.
Post-selected single-photon XPS.
  Most notably, for an average incident photon number of 0.5 (green-shaded region), the XPS for no-click and click events are $2\pm3$ and $-13\pm6\ \mu$rad, respectively, which definitively shows the effect of a single post-selected photon.
For the other data points, the average incident photon number and/or the signal center detuning is varied.
  Taking all the data points together, the magnitude of the post-selected single-photon XPS is $-18 \pm 4 \mu$rad.
  The inset shows the post-selected XPS versus $n_{inf} (2\pi\times18\ \textrm{MHz}) / \Delta_s$, inferred average photon number corrected for the variable signal detuning. 
  The solid line has a slope of $-14 \pm 1\ \mu$rad per photon. 
  Other relevant parameters include EIT window = 2 MHz and OD = 3.
  The data in the region shaded in blue are tests for systematics as explained in the text. 

\newpage
\textbf{Methods}


\emph{Atom preparation.}
The atoms are prepared in a magneto-optical trap using three beams and their retro-reflections along with a magnetic field gradient of 20 G/cm on the axis of the quadrupole coil.
Each of the three beams has a `trapping' (2.5 cm diameter) and a `repumper' (1 cm diameter) component.
The trapping beam is tuned to 20 MHz below the cycling transition $F=3\rightarrow F'=4$ and the repumper is tuned close to resonance on $F=2\rightarrow F'=3$ transition to keep the population in the trapping ground state. 
The beams and the magnetic field gradient are turned off every 22 ms, leaving the atoms to freely expand and allowing them to be probed for 1.5ms, as explained in figure \ref{fig_cycle}.

\emph{Probe phase measurement.}
The nonlinear effect we are interested in is a change in the probe refractive index, which is linear in the signal photon number and manifests itself as a phase shift of the probe field. 
In order to measure this probe phase shift, frequency-domain interferometry is used. 
The probe beam is comprised of two different frequency components, which copropagate through the atomic cloud.
One frequency component is tuned on resonance with the $F=2 \rightarrow F'=3$ transition while the other is +100 MHz detuned, serving as a phase reference; any phase or amplitude change due to the atoms appears as a phase or amplitude change of the resulting 100 MHz beating signal.
The probe beam is detected on a fast avalanche photodiode; demodulation of the resultant electrical signal at 100 MHz allows us to extract the phase and amplitude change experienced by the on-resonance probe field.
The analysis bandwidth for the demodulation is 2 MHz, which is matched to the EIT window for optimum signal-to-noise ratio.
A stable 10 MHz clock is used to generate a 100 MHz signal used as the reference for the modulation and demodulation. 
In order to eliminate the effects of any slow phase drifts, we use the average probe phase over 200 ns durations before and after the expected XPS as a phase background, which is subtracted off from the XPS. 
An important advantage of using frequency-domain interferometry and demodulation is that the measurement is insensitive to any source of variation that does not have a component at 100 MHz.
We measure a single-shot phase uncertainty of 50 mrad (there are approximately 4500 probe photons per measurement window, which is equivalent to a shot noise of 15 mrad).

\emph{Probe and coupling fields.}
Our master laser is locked 30MHz below the $F=2 \rightarrow F'=3$ transition of the D2 line in $^{85}$Rb.
A portion of this light is frequency-shifted using an acousto-optic modulator (AOM) driven at +130 MHz to produce the off-resonance component of the probe, which is then passed through a second AOM driven at -100 MHz to create the on-resonance component.
These two components are then combined on a beam splitter, forming the probe beat signal.
Another portion of the master laser is modulated using an electro-optic modulator (EOM) driven at 3 GHz.
This modulated light is used as the seed for injection-locking a diode laser; locking to the lower side-band of the modulation produces a coupling beam that is phase-locked to the probe, a necessary condition for establishing EIT.
This coupling beam passes through a third AOM, which serves to frequency shift it onto resonance with the $F=3 \rightarrow F'=3$ transition, as well as to shutter the coupling light on and off in sync with the atomic duty cycle.
The width of the EIT feature is linear in the coupling intensity and is measured to be 2MHz for all the data taken for this paper.
We use $\pi$ and $\sigma^+$ polarizations for the coupling and probe fields, respectively.

\emph{Signal pulses.}
A portion of the injection-locked laser light is split off to produce the signal pulses.
An AOM is used to frequency shift the light close to the $F=3\rightarrow F'=4$ transition as well as to create short pulses (40ns or 100ns FWHM). 
The temporal profiles of the signal pulses are measured on an avalanche photodiode.
By integrating the pulse power over time, the average energy, and therefore the average photon number, of each signal pulse can be measured.
The polarization of the signal light is $\sigma^+$.

\emph{Single-photon detection.}
The signal pulses are collected into a multimode fiber and are detected on an SPCM.
Other than the signal photons, any stray light which leaks into the signal collection can result in clicks.
We use gate signals with durations matching the signal pulses, to reject background photons.
The residual probability of background counts is $13\%$ for 100ns pulses and $6\%$ for 40ns signal pulses. 
Most of these result from the near-resonant scattering of the probe and coupling beams from the atoms.
The overall collection and detection efficiency is around $20\%$, and for higher incident photon numbers more attenuation is added to keep the total click rate around $20\%-30\%$.  
The details of the tagging procedure are explained in the Supplementary Information. 

In order to reduce the background photon rate for cases of incident average photon number of 0.5 and 1, the signal pulse duration was chosen to be 40 ns, while it had been 100 ns for the earlier data taken for higher signal power.
One might expect this change to make the signal pulses more intense and therefore make the nonlinear phase shift more than twice as big.
However, we have shown theoretically \cite{feizpour2014short} that because of the bandwidth mismatch between the EIT window (2 MHz here) and the signal pulses, the peak phase shift nearly saturates and we expect an enhancement of only 1.5. 
Therefore, given the size of our phase measurement uncertainty, the two values are expected not to be statistically different.

The inferred average photon number difference between click and no-click cases is slightly different from unity due to the effects of background counts, multiple-photon events and detection efficiency.
In order to calculate the nonlinear phase shift per post-selected single photon, the measured phase differences are divided by the values of the inferred average photon number difference.

\emph{Data collection.}
For each data point, we took approximately 300 million shots over 14 hours, 90 million of which resulted in clicks at the SPCM. 
Because of our tagging procedure 90 million shots were discarded, and out of the remaining shots we observed 60 million click events and 150 million  no-click events.

\emph{Focus size.}
The waists of the probe and signal beams inside the cloud are 13$\pm1\ \mu$m, corresponding to a (two-sided) Rayleigh range of roughly 1.4 mm.
The choice of this focus size is to ensure that the Rayleigh range of the beams matches the size of the cloud.
Focusing the beams tighter than the size of the cloud would  be detrimental because it would reduce the interaction length and increase the probe phase shot noise (the intensity of the probe is limited and fixed by the saturation intensity of atoms).
Also, focusing the beams any less tightly would produce a smaller intensity for given signal pulse energy, thereby decreases the size of the nonlinear effect.
 
\bibliographystyle{naturemag}
\bibliography{C:/Users/amir/Dropbox/PhD_thesis/etc/myThesisRefs}

\begin{thebibliography}{10}
\expandafter\ifx\csname url\endcsname\relax
  \def\url#1{\texttt{#1}}\fi
\expandafter\ifx\csname urlprefix\endcsname\relax\def\urlprefix{URL }\fi
\providecommand{\bibinfo}[2]{#2}
\providecommand{\eprint}[2][]{\url{#2}}

\bibitem{glauber1963coherent}
\bibinfo{author}{Glauber, R.~J.}
\newblock \bibinfo{title}{Coherent and incoherent states of the radiation
  field}.
\newblock \emph{\bibinfo{journal}{Physical Review}}
  \textbf{\bibinfo{volume}{131}}, \bibinfo{pages}{2766} (\bibinfo{year}{1963}).

\bibitem{glauber1963quantum}
\bibinfo{author}{Glauber, R.~J.}
\newblock \bibinfo{title}{The quantum theory of optical coherence}.
\newblock \emph{\bibinfo{journal}{Physical Review}}
  \textbf{\bibinfo{volume}{130}}, \bibinfo{pages}{2529} (\bibinfo{year}{1963}).

\bibitem{deutsch1992diphotons}
\bibinfo{author}{Deutsch, I.~H.}, \bibinfo{author}{Chiao, R.~Y.} \&
  \bibinfo{author}{Garrison, J.~C.}
\newblock \bibinfo{title}{Diphotons in a nonlinear fabry-perot resonator: Bound
  states of interacting photons in an optical ‘‘quantum wire’’}.
\newblock \emph{\bibinfo{journal}{Physical review letters}}
  \textbf{\bibinfo{volume}{69}}, \bibinfo{pages}{3627} (\bibinfo{year}{1992}).

\bibitem{firstenberg2013attractive}
\bibinfo{author}{Firstenberg, O.} \emph{et~al.}
\newblock \bibinfo{title}{Attractive photons in a quantum nonlinear medium}.
\newblock \emph{\bibinfo{journal}{Nature}} \textbf{\bibinfo{volume}{502}},
  \bibinfo{pages}{71--75} (\bibinfo{year}{2013}).
\newblock \urlprefix\url{http://www.nature.com/doifinder/10.1038/nature12512}.

\bibitem{ImotoQND}
\bibinfo{author}{Imoto, N.}, \bibinfo{author}{Haus, H.~A.} \&
  \bibinfo{author}{Yamamoto, Y.}
\newblock \bibinfo{title}{Quantum nondemolition measurement of the photon
  number via the optical kerr effect}.
\newblock \emph{\bibinfo{journal}{Phys. Rev. A}} \textbf{\bibinfo{volume}{32}},
  \bibinfo{pages}{2287--2292} (\bibinfo{year}{1985}).
\newblock \urlprefix\url{http://link.aps.org/doi/10.1103/PhysRevA.32.2287}.

\bibitem{braginsky_quantum_1980}
\bibinfo{author}{Braginsky, V.~B.}, \bibinfo{author}{Vorontsov, Y.~I.} \&
  \bibinfo{author}{Thorne, K.~S.}
\newblock \bibinfo{title}{Quantum nondemolition measurements}.
\newblock \emph{\bibinfo{journal}{Science}} \textbf{\bibinfo{volume}{209}},
  \bibinfo{pages}{547--557} (\bibinfo{year}{1980}).
\newblock \urlprefix\url{http://www.sciencemag.org/content/209/4456/547.short}.

\bibitem{grangier_quantum_1998}
\bibinfo{author}{Grangier, P.}, \bibinfo{author}{Levenson, J.~A.} \&
  \bibinfo{author}{Poizat, J.-P.}
\newblock \bibinfo{title}{Quantum non-demolition measurements in optics}.
\newblock \emph{\bibinfo{journal}{Nature}} \textbf{\bibinfo{volume}{396}},
  \bibinfo{pages}{537--542} (\bibinfo{year}{1998}).
\newblock
  \urlprefix\url{http://www.nature.com/nature/journal/v396/n6711/abs/396537a0.html}.

\bibitem{vitali2000complete}
\bibinfo{author}{Vitali, D.}, \bibinfo{author}{Fortunato, M.} \&
  \bibinfo{author}{Tombesi, P.}
\newblock \bibinfo{title}{Complete quantum teleportation with a kerr
  nonlinearity}.
\newblock \emph{\bibinfo{journal}{Physical review letters}}
  \textbf{\bibinfo{volume}{85}}, \bibinfo{pages}{445--448}
  (\bibinfo{year}{2000}).

\bibitem{HarrisSwitching1998}
\bibinfo{author}{Harris, S.~E.} \& \bibinfo{author}{Yamamoto, Y.}
\newblock \bibinfo{title}{Photon switching by quantum interference}.
\newblock \emph{\bibinfo{journal}{Phys. Rev. Lett.}}
  \textbf{\bibinfo{volume}{81}}, \bibinfo{pages}{3611--3614}
  (\bibinfo{year}{1998}).
\newblock \urlprefix\url{http://link.aps.org/doi/10.1103/PhysRevLett.81.3611}.

\bibitem{milburn1989fredkin}
\bibinfo{author}{Milburn, G.}
\newblock \bibinfo{title}{Quantum optical fredkin gate}.
\newblock \emph{\bibinfo{journal}{Physical Review Letters}}
  \textbf{\bibinfo{volume}{62}}, \bibinfo{pages}{2124} (\bibinfo{year}{1989}).

\bibitem{kimble1995cPhase}
\bibinfo{author}{Turchette, Q.~A.}, \bibinfo{author}{Hood, C.~J.},
  \bibinfo{author}{Lange, W.}, \bibinfo{author}{Mabuchi, H.} \&
  \bibinfo{author}{Kimble, H.~J.}
\newblock \bibinfo{title}{Measurement of conditional phase shifts for quantum
  logic}.
\newblock \emph{\bibinfo{journal}{Phys. Rev. Lett.}}
  \textbf{\bibinfo{volume}{75}}, \bibinfo{pages}{4710--4713}
  (\bibinfo{year}{1995}).
\newblock \urlprefix\url{http://link.aps.org/doi/10.1103/PhysRevLett.75.4710}.

\bibitem{raimond_manipulating_2001}
\bibinfo{author}{Raimond, J.-M.}, \bibinfo{author}{Brune, M.} \&
  \bibinfo{author}{Haroche, S.}
\newblock \bibinfo{title}{Manipulating quantum entanglement with atoms and
  photons in a cavity}.
\newblock \emph{\bibinfo{journal}{Reviews of Modern Physics}}
  \textbf{\bibinfo{volume}{73}}, \bibinfo{pages}{565} (\bibinfo{year}{2001}).
\newblock
  \urlprefix\url{http://journals.aps.org/rmp/abstract/10.1103/RevModPhys.73.565}.

\bibitem{guerlin_progressive_2007}
\bibinfo{author}{Guerlin, C.} \emph{et~al.}
\newblock \bibinfo{title}{Progressive field-state collapse and quantum
  non-demolition photon counting}.
\newblock \emph{\bibinfo{journal}{Nature}} \textbf{\bibinfo{volume}{448}},
  \bibinfo{pages}{889--893} (\bibinfo{year}{2007}).
\newblock
  \urlprefix\url{http://www.nature.com/nature/journal/v448/n7156/abs/nature06057.html}.

\bibitem{gleyzes_quantum_2007}
\bibinfo{author}{Gleyzes, S.} \emph{et~al.}
\newblock \bibinfo{title}{Quantum jumps of light recording the birth and death
  of a photon in a cavity}.
\newblock \emph{\bibinfo{journal}{Nature}} \textbf{\bibinfo{volume}{446}},
  \bibinfo{pages}{297--300} (\bibinfo{year}{2007}).
\newblock
  \urlprefix\url{http://www.nature.com/nature/journal/v446/n7133/abs/nature05589.html}.

\bibitem{rauschenbeutel_step-by-step_2000}
\bibinfo{author}{Rauschenbeutel, A.} \emph{et~al.}
\newblock \bibinfo{title}{Step-by-step engineered multiparticle entanglement}.
\newblock \emph{\bibinfo{journal}{Science}} \textbf{\bibinfo{volume}{288}},
  \bibinfo{pages}{2024--2028} (\bibinfo{year}{2000}).
\newblock
  \urlprefix\url{http://www.sciencemag.org/content/288/5473/2024.short}.

\bibitem{wallraff2004strong}
\bibinfo{author}{Wallraff, A.} \emph{et~al.}
\newblock \bibinfo{title}{Strong coupling of a single photon to a
  superconducting qubit using circuit quantum electrodynamics}.
\newblock \emph{\bibinfo{journal}{Nature}} \textbf{\bibinfo{volume}{431}},
  \bibinfo{pages}{162--167} (\bibinfo{year}{2004}).

\bibitem{devoret2013superconducting}
\bibinfo{author}{Devoret, M.} \& \bibinfo{author}{Schoelkopf, R.}
\newblock \bibinfo{title}{Superconducting circuits for quantum information: an
  outlook}.
\newblock \emph{\bibinfo{journal}{Science}} \textbf{\bibinfo{volume}{339}},
  \bibinfo{pages}{1169--1174} (\bibinfo{year}{2013}).

\bibitem{schmidt1996giant}
\bibinfo{author}{Schmidt, H.} \& \bibinfo{author}{Imamoglu, A.}
\newblock \bibinfo{title}{Giant kerr nonlinearities obtained by
  electromagnetically induced transparency}.
\newblock \emph{\bibinfo{journal}{Optics letters}}
  \textbf{\bibinfo{volume}{21}}, \bibinfo{pages}{1936--1938}
  (\bibinfo{year}{1996}).

\bibitem{matsuda2009observation}
\bibinfo{author}{Matsuda, N.}, \bibinfo{author}{Shimizu, R.},
  \bibinfo{author}{Mitsumori, Y.}, \bibinfo{author}{Kosaka, H.} \&
  \bibinfo{author}{Edamatsu, K.}
\newblock \bibinfo{title}{Observation of optical-fibre kerr nonlinearity at the
  single-photon level}.
\newblock \emph{\bibinfo{journal}{Nature photonics}}
  \textbf{\bibinfo{volume}{3}}, \bibinfo{pages}{95--98} (\bibinfo{year}{2009}).

\bibitem{venkataraman2013phase}
\bibinfo{author}{Venkataraman, V.}, \bibinfo{author}{Saha, K.} \&
  \bibinfo{author}{Gaeta, A.~L.}
\newblock \bibinfo{title}{Phase modulation at the few-photon level for
  weak-nonlinearity-based quantum computing}.
\newblock \emph{\bibinfo{journal}{Nature Photonics}}
  \textbf{\bibinfo{volume}{7}}, \bibinfo{pages}{138--141}
  (\bibinfo{year}{2013}).

\bibitem{PhysRevLett.111.193601}
\bibinfo{author}{O'Shea, D.}, \bibinfo{author}{Junge, C.},
  \bibinfo{author}{Volz, J.} \& \bibinfo{author}{Rauschenbeutel, A.}
\newblock \bibinfo{title}{Fiber-optical switch controlled by a single atom}.
\newblock \emph{\bibinfo{journal}{Phys. Rev. Lett.}}
  \textbf{\bibinfo{volume}{111}}, \bibinfo{pages}{193601}
  (\bibinfo{year}{2013}).
\newblock
  \urlprefix\url{http://link.aps.org/doi/10.1103/PhysRevLett.111.193601}.

\bibitem{hendrickson2009nonlinear}
\bibinfo{author}{Hendrickson, S.}, \bibinfo{author}{Pittman, T.} \&
  \bibinfo{author}{Franson, J.}
\newblock \bibinfo{title}{Nonlinear transmission through a tapered fiber in
  rubidium vapor}.
\newblock \emph{\bibinfo{journal}{JOSA B}} \textbf{\bibinfo{volume}{26}},
  \bibinfo{pages}{267--271} (\bibinfo{year}{2009}).

\bibitem{spillane2008observation}
\bibinfo{author}{Spillane, S.} \emph{et~al.}
\newblock \bibinfo{title}{Observation of nonlinear optical interactions of
  ultralow levels of light in a tapered optical nanofiber embedded in a hot
  rubidium vapor}.
\newblock \emph{\bibinfo{journal}{Physical review letters}}
  \textbf{\bibinfo{volume}{100}}, \bibinfo{pages}{233602}
  (\bibinfo{year}{2008}).

\bibitem{volz_nonlinear_2014}
\bibinfo{author}{Volz, J.}, \bibinfo{author}{Scheucher, M.},
  \bibinfo{author}{Junge, C.} \& \bibinfo{author}{Rauschenbeutel, A.}
\newblock \bibinfo{title}{Nonlinear π phase shift for single fibre-guided
  photons interacting with a single resonator-enhanced atom}.
\newblock \emph{\bibinfo{journal}{Nature Photonics}}
  \textbf{\bibinfo{volume}{advance online publication}} (\bibinfo{year}{2014}).
\newblock
  \urlprefix\url{http://www.nature.com.myaccess.library.utoronto.ca/nphoton/journal/vaop/ncurrent/full/nphoton.2014.253.html}.

\bibitem{urban2009observation}
\bibinfo{author}{Urban, E.} \emph{et~al.}
\newblock \bibinfo{title}{Observation of rydberg blockade between two atoms}.
\newblock \emph{\bibinfo{journal}{Nature Physics}}
  \textbf{\bibinfo{volume}{5}}, \bibinfo{pages}{110--114}
  (\bibinfo{year}{2009}).

\bibitem{saffman_quantum_2010}
\bibinfo{author}{Saffman, M.}, \bibinfo{author}{Walker, T.~G.} \&
  \bibinfo{author}{Mølmer, K.}
\newblock \bibinfo{title}{Quantum information with rydberg atoms}.
\newblock \emph{\bibinfo{journal}{Reviews of Modern Physics}}
  \textbf{\bibinfo{volume}{82}}, \bibinfo{pages}{2313--2363}
  (\bibinfo{year}{2010}).
\newblock \urlprefix\url{http://link.aps.org/doi/10.1103/RevModPhys.82.2313}.

\bibitem{parigi_observation_2012}
\bibinfo{author}{Parigi, V.} \emph{et~al.}
\newblock \bibinfo{title}{Observation and measurement of interaction-induced
  dispersive optical nonlinearities in an ensemble of cold rydberg atoms}.
\newblock \emph{\bibinfo{journal}{Physical review letters}}
  \textbf{\bibinfo{volume}{109}}, \bibinfo{pages}{233602}
  (\bibinfo{year}{2012}).
\newblock
  \urlprefix\url{http://journals.aps.org/prl/abstract/10.1103/PhysRevLett.109.233602}.

\bibitem{peyronel2012quantum}
\bibinfo{author}{Peyronel, T.} \emph{et~al.}
\newblock \bibinfo{title}{Quantum nonlinear optics with single photons enabled
  by strongly interacting atoms}.
\newblock \emph{\bibinfo{journal}{Nature}} \textbf{\bibinfo{volume}{488}},
  \bibinfo{pages}{57--60} (\bibinfo{year}{2012}).

\bibitem{Chen16082013}
\bibinfo{author}{Chen, W.} \emph{et~al.}
\newblock \bibinfo{title}{All-optical switch and transistor gated by one stored
  photon}.
\newblock \emph{\bibinfo{journal}{Science}} \textbf{\bibinfo{volume}{341}},
  \bibinfo{pages}{768--770} (\bibinfo{year}{2013}).
\newblock
  \urlprefix\url{http://www.sciencemag.org/content/341/6147/768.abstract}.
\newblock \eprint{http://www.sciencemag.org/content/341/6147/768.full.pdf}.

\bibitem{PhysRevLett.112.073901}
\bibinfo{author}{Baur, S.}, \bibinfo{author}{Tiarks, D.},
  \bibinfo{author}{Rempe, G.} \& \bibinfo{author}{D\"urr, S.}
\newblock \bibinfo{title}{Single-photon switch based on rydberg blockade}.
\newblock \emph{\bibinfo{journal}{Phys. Rev. Lett.}}
  \textbf{\bibinfo{volume}{112}}, \bibinfo{pages}{073901}
  (\bibinfo{year}{2014}).
\newblock
  \urlprefix\url{http://link.aps.org/doi/10.1103/PhysRevLett.112.073901}.

\bibitem{guerreiro_nonlinear_2014}
\bibinfo{author}{Guerreiro, T.} \emph{et~al.}
\newblock \bibinfo{title}{Nonlinear interaction between single photons}.
\newblock \emph{\bibinfo{journal}{Physical Review Letters}}
  \textbf{\bibinfo{volume}{113}}, \bibinfo{pages}{173601}
  (\bibinfo{year}{2014}).
\newblock
  \urlprefix\url{http://link.aps.org/doi/10.1103/PhysRevLett.113.173601}.

\bibitem{shapiroPRA06single}
\bibinfo{author}{Shapiro, J.~H.}
\newblock \bibinfo{title}{Single-photon kerr nonlinearities do not help quantum
  computation}.
\newblock \emph{\bibinfo{journal}{Phys. Rev. A}} \textbf{\bibinfo{volume}{73}},
  \bibinfo{pages}{062305} (\bibinfo{year}{2006}).
\newblock \urlprefix\url{http://link.aps.org/doi/10.1103/PhysRevA.73.062305}.

\bibitem{shapiroNJP07weak}
\bibinfo{author}{Shapiro, J.~H.} \& \bibinfo{author}{Razavi, M.}
\newblock \bibinfo{title}{Continuous-time cross-phase modulation and quantum
  computation}.
\newblock \emph{\bibinfo{journal}{New Journal of Physics}}
  \textbf{\bibinfo{volume}{9}}, \bibinfo{pages}{16} (\bibinfo{year}{2007}).

\bibitem{GeaBanaclochePRA10impossiblity}
\bibinfo{author}{Gea-Banacloche, J.}
\newblock \bibinfo{title}{Impossibility of large phase shifts via the giant
  kerr effect with single-photon wave packets}.
\newblock \emph{\bibinfo{journal}{Phys. Rev. A}} \textbf{\bibinfo{volume}{81}},
  \bibinfo{pages}{043823} (\bibinfo{year}{2010}).
\newblock \urlprefix\url{http://link.aps.org/doi/10.1103/PhysRevA.81.043823}.

\bibitem{nemotoPRL04nearly}
\bibinfo{author}{Nemoto, K.} \& \bibinfo{author}{Munro, W.~J.}
\newblock \bibinfo{title}{Nearly deterministic linear optical controlled-not
  gate}.
\newblock \emph{\bibinfo{journal}{Physical review letters}}
  \textbf{\bibinfo{volume}{93}}, \bibinfo{pages}{250502}
  (\bibinfo{year}{2004}).

\bibitem{munroNJP05weak}
\bibinfo{author}{Munro, W.~J.}, \bibinfo{author}{Nemoto, K.} \&
  \bibinfo{author}{Spiller, T.~P.}
\newblock \bibinfo{title}{Weak nonlinearities: a new route to optical quantum
  computation}.
\newblock \emph{\bibinfo{journal}{New Journal of Physics}}
  \textbf{\bibinfo{volume}{7}}, \bibinfo{pages}{137} (\bibinfo{year}{2005}).

\bibitem{loPRAattojoulePhysRevA.83.041804}
\bibinfo{author}{Lo, H.-Y.} \emph{et~al.}
\newblock \bibinfo{title}{Electromagnetically-induced-transparency-based
  cross-phase-modulation at attojoule levels}.
\newblock \emph{\bibinfo{journal}{Phys. Rev. A}} \textbf{\bibinfo{volume}{83}},
  \bibinfo{pages}{041804} (\bibinfo{year}{2011}).

\bibitem{smith_capture_2002}
\bibinfo{author}{Smith, W.~P.}, \bibinfo{author}{Reiner, J.~E.},
  \bibinfo{author}{Orozco, L.~A.}, \bibinfo{author}{Kuhr, S.} \&
  \bibinfo{author}{Wiseman, H.~M.}
\newblock \bibinfo{title}{Capture and release of a conditional state of a
  cavity {QED} system by quantum feedback}.
\newblock \emph{\bibinfo{journal}{Physical Review Letters}}
  \textbf{\bibinfo{volume}{89}}, \bibinfo{pages}{133601}
  (\bibinfo{year}{2002}).
\newblock
  \urlprefix\url{http://link.aps.org/doi/10.1103/PhysRevLett.89.133601}.

\bibitem{foster_third-order_2002}
\bibinfo{author}{Foster, G.~T.}, \bibinfo{author}{Smith, W.~P.},
  \bibinfo{author}{Reiner, J.~E.} \& \bibinfo{author}{Orozco, L.~A.}
\newblock \bibinfo{title}{Third-order correlations in cavity quantum
  electrodynamics}.
\newblock \emph{\bibinfo{journal}{Journal of Optics B: Quantum and
  Semiclassical Optics}} \textbf{\bibinfo{volume}{4}}, \bibinfo{pages}{S281}
  (\bibinfo{year}{2002}).
\newblock \urlprefix\url{http://iopscience.iop.org/1464-4266/4/4/306}.

\bibitem{foster_time-dependent_2002}
\bibinfo{author}{Foster, G.~T.}, \bibinfo{author}{Smith, W.~P.},
  \bibinfo{author}{Reiner, J.~E.} \& \bibinfo{author}{Orozco, L.~A.}
\newblock \bibinfo{title}{Time-dependent electric field fluctuations at the
  subphoton level}.
\newblock \emph{\bibinfo{journal}{Physical Review A}}
  \textbf{\bibinfo{volume}{66}}, \bibinfo{pages}{033807}
  (\bibinfo{year}{2002}).
\newblock \urlprefix\url{http://link.aps.org/doi/10.1103/PhysRevA.66.033807}.

\bibitem{fleischhauer2005electromagnetically}
\bibinfo{author}{Fleischhauer, M.}, \bibinfo{author}{Imamoglu, A.} \&
  \bibinfo{author}{Marangos, J.}
\newblock \bibinfo{title}{Electromagnetically induced transparency: Optics in
  coherent media}.
\newblock \emph{\bibinfo{journal}{Reviews of Modern Physics}}
  \textbf{\bibinfo{volume}{77}}, \bibinfo{pages}{633} (\bibinfo{year}{2005}).

\bibitem{feizpour2014short}
\bibinfo{author}{Feizpour, A.}, \bibinfo{author}{Dmochowski, G.} \&
  \bibinfo{author}{Steinberg, A.~M.}
\newblock \bibinfo{title}{Short-pulse cross-phase modulation in an
  electromagnetically-induced-transparency medium}.
\newblock \emph{\bibinfo{journal}{arXiv preprint arXiv:1406.0245}}
  (\bibinfo{year}{2014}).

\bibitem{PRL_HarrisHau}
\bibinfo{author}{Harris, S.~E.} \& \bibinfo{author}{Hau, L.~V.}
\newblock \bibinfo{title}{Nonlinear optics at low light levels}.
\newblock \emph{\bibinfo{journal}{Phys. Rev. Lett.}}
  \textbf{\bibinfo{volume}{82}}, \bibinfo{pages}{4611--4614}
  (\bibinfo{year}{1999}).
\newblock \urlprefix\url{http://link.aps.org/doi/10.1103/PhysRevLett.82.4611}.

\end{thebibliography}

\newpage
\textbf{Supplementary Information} is available in the online version of the paper.

\textbf{Acknowledgements} 
This work was funded by NSERC, CIFAR, and QuantumWorks. We would like to thank Alex Hayat for useful discussions, and Alan Stummer for designing and building several electronic devices for this experiment.

\textbf{Author Contributions}
All authors have contributed to the design of the experiment, interpretation of the results, preparation and revisions of the manuscript.

\textbf{Author Information}
Reprints and permissions information is available at
www.nature.com/reprints. The authors declare no competing financial interests.
Readers are welcome to comment on the online version of the paper. Correspondence
should be addressed to A.F. (feizpour@physics.utoronto.ca)

\appendix
\newpage
\section{Supplementary information}
\label{sec_infPhNum}

\emph{Inferred average photon number in the interaction region}.
In order to calculate the most probable number of signal photons interacting with the probe, one can use Bayesian inference to obtain the conditional probability of having had $n_{ph}$ photons in the interaction region given the occurrence (``yes") or lack (``no") of a signal photon detection event (`click') after the interaction:

\begin{eqnarray}
P(n_{ph}|\textrm{no}) &=& \frac{1}{N_0}P(\textrm{no}|n_{ph})P(n_{ph}) = e^{-(1-\eta)|\alpha|^2}\frac{|\alpha|^{2n_{ph}}}{n_{ph}!}(1-\eta)^{n_{ph}}\nonumber\\ 
P(n_{ph}|\textrm{yes}) &=& \frac{1}{N_1}P(\textrm{yes}|n_{ph})P(n_{ph}) 
= \frac{e^{-|\alpha|^2}}{1-e^{-\eta|\alpha|^2}}\frac{|\alpha|^{2n_{ph}}}{n_{ph}!}(1-(1-\eta)^{n_{ph}}),
\end{eqnarray}

\noindent
where $N_0$ and $N_1$ are normalization factors, $\eta$ is the overall collection and detection efficiency, and $P(n_{ph}) = \exp(-|\alpha|^2) |\alpha|^{2n_{ph}} / n_{ph}!$ is the incident signal photon number distribution with average $|\alpha|^2$.
Using the conditional probabilities given above, one can calculate the average photon number in the interaction region in each case,

\begin{eqnarray}
\bar{n}_{inf,\textrm{\scriptsize{no}}} & = & |\alpha|^2 (1-\eta), \nonumber\\
\bar{n}_{inf,\textrm{\scriptsize{yes}}} & = & |\alpha|^2 \left(1+\eta \frac{e^{-\eta|\alpha|^2}}{1-e^{-\eta|\alpha|^2}} \right)\nonumber\\
&=& \bar{n}_{inf,\textrm{\scriptsize{no}}} + \frac{\eta |\alpha|^2}{P_{sig}(\textrm{yes})},
\label{eq_n_inff}
\end{eqnarray}

\noindent
where $P_{sig}(\textrm{yes}) = 1 - P_{sig}(\textrm{no}) = 1 - e^{-\eta|\alpha|^2}$ is the probability of a detection event.
For low count rates, $\eta |\alpha|^2\ll 1$, the expression above reduces to $\bar{n}_{inf,\textrm{\scriptsize{yes}}} \approx \bar{n}_{inf,\textrm{\scriptsize{no}}} +  1$; the difference in the inferred average photon number between click and no-click events is unity, independent of both the detection efficiency and the average incident photon number.
It is straightforward to include the effect of background photons:
$P_{sig}(\textrm{no}) \rightarrow P_{sig}(\textrm{no})P_{bkg}(\textrm{no})$ and 
$P_{sig}(\textrm{yes}) \rightarrow 1 - P_{sig}(\textrm{no})P_{bkg}(\textrm{no})$ where $P_{bkg}(\textrm{no})$ is the probability of getting \emph{no} clicks from background photons.
 
\begin{figure}[h]
  \centering %
  \includegraphics[width = 0.5\columnwidth]{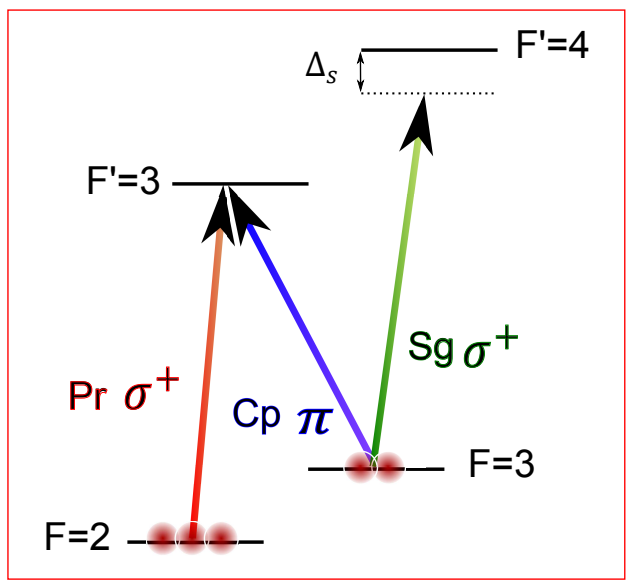}
  \caption{Level scheme.   
  \label{fig_levelScheme}}
\end{figure}

\emph{Detuning-dependence of the cross-phase shift}. 
Figure \ref{fig_levelScheme} shows the level scheme used for this experiment; the presence of an off-resonant signal pulse introduces an ac-Stark shift, which effectively detunes the probe field from resonance causing it to acquire a phase shift.
Here, we are interested in the dependence of this phase shift on the detuning of the signal pulse.
So long as the probe samples the linear portion of its refractive index profile, the detuning-dependence of the nonlinear phase shift will be the same as the detuning dependence of the ac-Stark shift itself.
In the limit of weak signal pulses and weak probe field, the ac-Stark shift as a function of signal detuning, $\Delta_s$, is given by,

\begin{equation}
	\Delta_{ACS} = \frac{-\Omega_s^2\Delta_s}{\Delta_s^2 + (\Gamma/2)^2},
\end{equation}

\noindent
where $\Gamma=2\pi\times6$ MHz is the excited state linewidth and $\Omega_s^2$ is proportional to the intensity (and therefore the photon number) of the signal pulse.
This detuning dependence has a dispersion-like shape that goes to zero on resonance and reaches extrema at one-half of the line-width, $\Gamma$.

A second contribution to the detuning-dependence of the nonlinear phase shift arises from the non-vanishing atomic population in the ground state addressed by the coupling field.
The population in each ground state of an EIT system is determined by the ratio of the coupling and probe field intensities.
In our case, the probe intensity is sufficient to bring some population into the ground state addressed by the coupling and signal fields, which leads to finite signal absorption; see figure \ref{fig_levelScheme}.
This leads to an effective number of signal photons in the interaction region, 

\begin{equation}
N_{eff} = N_0\frac{1-e^{-d_s(\Delta_s)}}{d_s(\Delta_s)}
\end{equation}

\noindent where $d_s(\Delta_s) = d_0\Gamma^2/(4\Delta_s^2 + \Gamma^2) $ is the optical density of the signal transition at the detuning $\Delta_s$ and $N_0$ is the incident photon number.
Here $d_0$ is the on-resonance optical density.
The nonlinear phase shift is then proportional to the product of these two contributions,

\begin{equation}
\phi(\Delta_s) = - 2\phi_m \frac{\Delta_s\Gamma/2}{\Delta_s^2 + (\Gamma/2)^2} \frac{1 - e^{-d_s(\Delta_s)}}{d_s(\Delta_s)},
\end{equation}

\noindent
where $\phi_m$ is a proportionality constant. 

\begin{figure}[h]
  \centering %
  \includegraphics[width = \columnwidth]{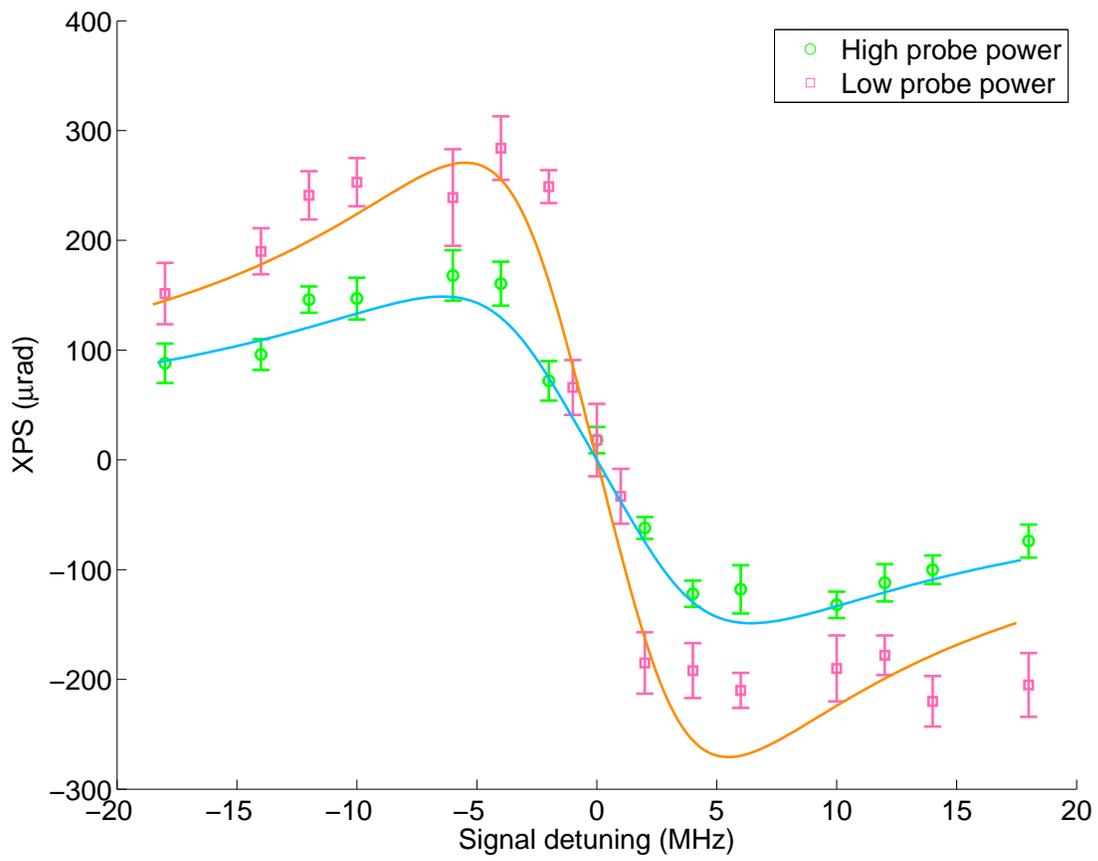}
  \caption{XPS versus signal detuning.   
  \label{fig_vsDetuning}}
\end{figure}

Figure \ref{fig_vsDetuning} shows the measured nonlinear phase shift as a function of signal detuning, along with fits based on all these contributions.
The detuning dependence is measured for two different probe intensities, keeping the coupling intensity fixed.
The fit parameters are $d_0 = 4\pm2$, $\phi_m = 500\pm100\ \mu$rad for low probe power, and $d_0 = 5\pm2$, $\phi_m = 300\pm40\ \mu$rad for high probe power.
It can be seen that in the case of higher probe power the nonlinear phase shift is smaller because of the higher signal absorption.

\emph{Theoretical value of the cross-phase shift.}
Previously, we showed theoretically that the temporal profile of the XPS expected for a single-photon Gaussian signal pulse with an rms duration of $\tau_s$ interacting with an EIT medium with a response time of $\tau$ is given by \cite{feizpour2014short}

\begin{eqnarray}
\phi(t) = \frac{\phi_0}{2\tau} e^{\tau_s^2/2\tau^2} \exp(-t/\tau) \left(1 + \textrm{erf}(t/\sqrt{2}\tau_s - \tau_s/\sqrt{2}\tau)\right)
\label{phiOfTee}
\end{eqnarray}

\noindent where $\textrm{erf}(x) = 2/\sqrt{\pi} \int_0^x dx' \exp(-x'^2)$ is the error function,  and

\begin{equation}
\phi_0 = \frac{\Gamma}{-4\Delta_s} \frac{\sigma_{at}}{\pi w_0^2} \frac{d}{\Delta_{EIT}}
\label{eq_phi_int}
\end{equation}

\noindent
is the integrated XPS per signal photon.
Here, $\Gamma$ is the excited state linewidth, $\Delta_s$ is the signal detuning, $\sigma_{at}$ is the atomic cross section as for the signal, $w_0$ is the beam waist, $d$ is the change in the optical density for the on-resonance probe without and with a resonant coupling beam, and $\Delta_{EIT}$ is the full-width at half-maximum of the transparency.
For our experimental parameters ($\Gamma = 2\pi\times 6$ MHz, $|\Delta_s| = 2\pi\times 18$ MHz, $A/\sigma_{at} = 3000$, $d=2$, $\Delta_{EIT} = 2\pi\times 2$ MHz, $\tau_s = 40$ ns, and $\tau = 250$ ns) the XPS has a peak equal to 13 $\mu$rad.

The maximum achievable cross-phase shift per photon in the N-scheme, due to group velocity mismatch issues, is

\begin{equation}
|\phi_{max}| = \frac{\Gamma}{4|\Delta_s|} \frac{\sigma_{at}}{A},
\end{equation}

\noindent as shown by Harris and Hau \cite{PRL_HarrisHau}.
For the parameters of our experiment this value is $28 \ \mu$rad.
The phase shift we measure is lower than this value because our optical density is not high enough to saturate the limit posed by group velocity mismatch. 

\begin{figure}[h]
  \centering %
  \includegraphics[width = \columnwidth]{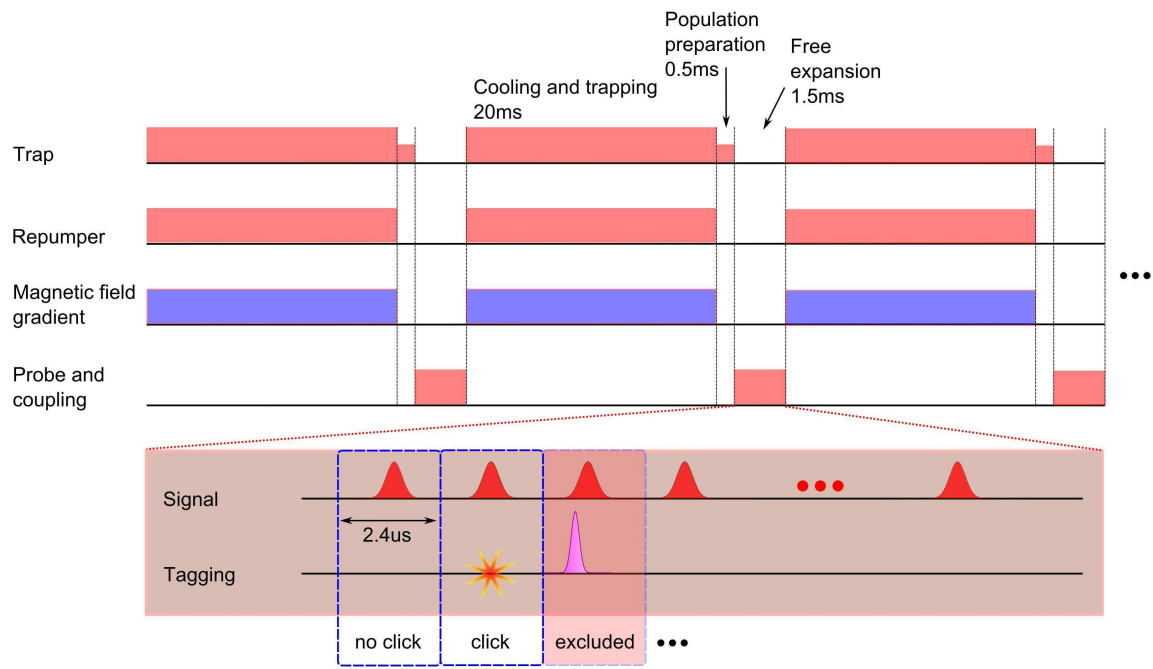}
  \caption{Measurement and atom cycle. 
  \label{fig_cycle}}
\end{figure}

\emph{Measurement and atom cycle.}
The atoms are first captured for 20 ms in the $F=3$ ground state using a magneto-optical trap (MOT) with a trapping beam red-detuned by $20$MHz from the $F=3 \rightarrow F'=4$ cycling transition of the D2 line in $^85$Rb.
Then, for 0.5 ms, this trapping light is tuned closer to the $F=3 \rightarrow F'=3$ transition so as to prepare the population in the $F=2$ ground state.
During this time, the magnetic field gradient is turned off (and kept off until the next recapture period) in order to avoid spatially varying Zeeman shifts, which would lead to dephasing of the EIT system.
After this population preparation stage, all MOT beams are turned off, leaving the cloud to freely expand while the probe and coupling fields are turned on for 1.5 ms.
The amplitude and phase of the probe field is continuously measured during this period.
Short pulses of signal light (40 ns or 100 ns FWHM duration) are sent in every 2.4 $\mu$s, each pulse constituting one `shot'. 
After 1.5 ms, the probe, coupling and signal fields are turned off and the recapturing period begins again.
The cross-phase shift that the signal pulses write on the probe field is obtained by splicing the 1.5 ms trace of probe phase data into 2.4 $\mu$s shots and averaging over many runs; see figure \ref{fig_cycle}.


\emph{Tagging procedure.}
In order to see the effect of a post-selected single photon, we need to select out the shots that lead to the detection of a signal photon after the interaction.
This is achieved by sending a bright flash of light into the probe detector conditioned on the click from the single-photon counting module (SPCM) that is used to detect these signal photons.
This flash of light appears as a spike in the amplitude of the probe field and allows us to pick out the shots that lead to a single photon detection event.

In order to exclude false positives arising from background counts of the SPCM, we use time-gating to accept only those clicks which occurred during the signal pulses themselves.
A logical AND gate is used to do the time-gating, the output of which triggers the flash of light that enters the probe detector.

Prior to the interaction, signal pulses pass through a polarizing beam splitter (PBS). 
One output of the PBS is further attenuated and sent to the atom cloud, serving as the single-photon-level signal pulses described above.
The other output of the PBS is directed to a fast photo-diode, the electrical output of which goes to one input of the AND gate.
The TTL output from the SPCM serves as the second input to the AND gate and in this way we can exclude clicks that arise from background or dark counts of the SPCM.

Recalling that our probe detection scheme involves demodulation of the probe signal at 100 MHz, the flash of light used for tagging must contain a 100 MHz frequency component in order to make it through the demodulation.
This is achieved by using the output of the logical AND gate to switch on an AOM, which is driven at 100 MHz. 
Picking off the first and second orders of light passing through this AOM produces a 100 MHz beat signal that is then sent into the probe detector.
The AOM driver is switched on for 200 ns, producing short bursts of light, which show up as spikes in the probe field amplitude.
The output of the logical AND gate is delayed so that these spikes arrive at the probe detector during the subsequent shot.
For this reason, we exclude all shots that follow a successful detection event.
An important technical advantage of this method to select out shots containing single photons is that it is electrically decoupled from the phase measurement electronics to avoid any cross-talk between the two systems. 

%
%
%
%

\newpage
\textbf{Extended data}

\emph{Figure \ref{fig_levelScheme}}.
Level scheme.
  The level scheme used to observe cross-phase modulation using the D2 line of $^{85}$Rb atoms. 
  The ac-Stark shift due to the signal pulses, pulls the probe out of the EIT condition and this appears as a refractive index change proportional to the signal intensity.

\begin{figure}[h]
  \centering %
  \includegraphics[width = \columnwidth]{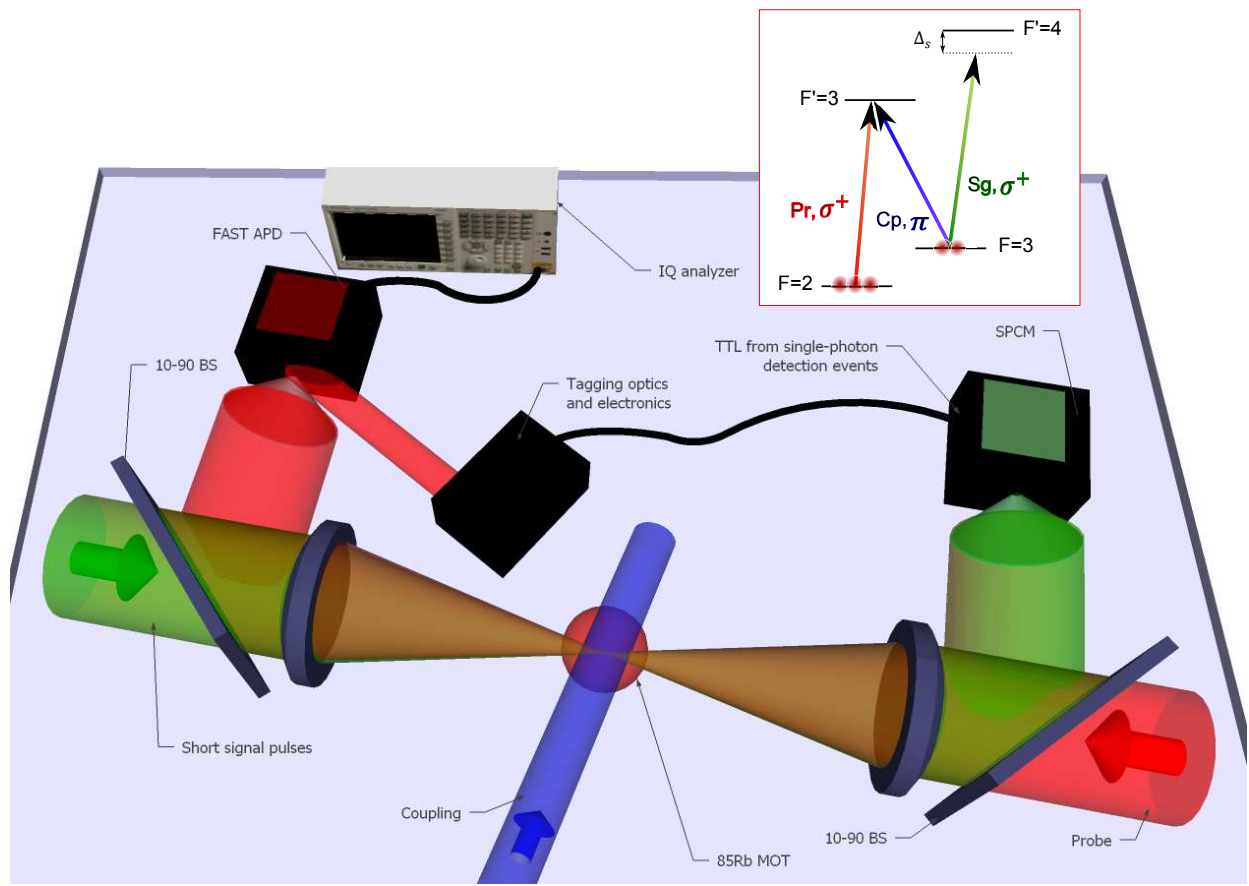}
  \caption{Schematic of the experimental setup. 
  \label{fig_setup}}
\end{figure}

\emph{Figure \ref{fig_vsDetuning}}.
XPS versus signal detuning.
  The nonlinear phase shift is caused by the ac-Stark shift due to the signal pulses.
  Therefore, it has the same dependence on signal detuning as the ac-Stark shift.
  This scaling also depends on probe power because more probe power results in a larger population in $F=3$ ground state (see the inset of fig. \ref{fig_setup}) which means a larger signal absorption. 
  The overall effect is broadening and smearing of the dispersion-like scaling at higher probe powers.
  
\emph{Figure \ref{fig_cycle}}.
Measurement and atom cycle. 
The $^{85}$Rb atoms are prepared in a magneto-optical trap using a cycle as explained in the text.
During the free expansion time, signal pulses are sent into the interaction region every 2.4 $\mu s$, one `shot', and the phase of the probe is monitored continuously.
The interesting quantity here is the difference in the phase of the probe for cases that there is a signal single-photon detection, `click', compared to the no-click cases. 

\newpage

\newpage

\newpage

\newpage

\newpage

\newpage
	
\newpage

\end{document}